\documentclass[runningheads]{llncs}
\usepackage[T1]{fontenc}
\usepackage{amssymb}
\usepackage{graphicx}

\usepackage{booktabs}
\usepackage{url}
\usepackage{algpseudocode}
\usepackage{amsmath}
\usepackage{amsfonts}
\usepackage[ruled,linesnumbered]{algorithm2e}
\usepackage{multicol}     
\usepackage{multirow}
\usepackage[misc]{ifsym}
\usepackage{color}
\begin{document}
\title{Phishing Fraud Detection on Ethereum using Graph Neural Network}
%
%

\author{Panpan Li\inst{1,2} \and
Yunyi Xie\inst{1,2} \and
Xinyao Xu\inst{1,2} \and    \\
Jiajun Zhou\inst{1,2} \textsuperscript{(\Letter)}     \and
Qi Xuan\inst{1,2}
}
\authorrunning{P. Li et al.}
%
\institute{Institute of Cyberspace Security, Zhejiang University of Technology, \\Hangzhou 310023, China  \and
           College of Information Engineering, Zhejiang University of Technology, \\Hangzhou 310023, China\\
\email{jjzhou@zjut.edu.cn}}
\maketitle              

\begin{abstract}
Blockchain has widespread applications in the financial field but has also attracted increasing cybercrimes. 
Recently, phishing fraud has emerged as a major threat to blockchain security, calling for the development of effective regulatory strategies. 
Nowadays network science has been widely used in modeling Ethereum transaction data, further introducing the network representation learning technology to analyze the transaction patterns.
In this paper, we consider phishing detection as a graph classification task and propose an end-to-end \textbf{P}hishing \textbf{D}etection \textbf{G}raph \textbf{N}eural \textbf{N}etwork framework (PDGNN). 
Specifically, we first construct a lightweight Ethereum transaction network and extract transaction subgraphs of collected phishing accounts. 
Then we propose an end-to-end detection model based on Chebyshev-GCN to precisely distinguish between normal and phishing accounts. 
Extensive experiments on five Ethereum datasets demonstrate that our PDGNN significantly outperforms general phishing detection methods and scales well in large transaction networks.

\end{abstract}
\section{Introduction}
Thanks to the development of blockchain technology, the past few years have witnessed the emergence of more than 1,100 cryptocurrencies in the financial field, further reforming the financial system.
In general, blockchain technology can be described as a distributed ledger maintained by a peer-to-peer network through a consensus mechanism. 
Its non-intermediary trust system changes the traditional mechanism of establishing and maintaining trust-based Internet central institutions, allowing accounts to freely exchange cryptocurrencies or other fiat currencies without relying on traditional third parties.

However, the financial nature of cryptocurrencies makes them the target of various financial crimes~\cite{holub2018coinhoarder}. 
According to CoinGecko data, the current market value of cryptocurrencies has risen to 1.84 trillion, and the 24-hour trading volume is 78.687 billion.
Moreover, the anonymization and weak regulation of blockchain platforms also exacerbate the problem.
According to incomplete statistics, in the first half of 2017 alone, 30,287 users suffered financial fraud on the Ethereum platform, including phishing scams, Ponzi schemes, and ransomware, with a total economic loss of 225 million. 
Among these frauds, over 50$\%$ were classified as phishing scams targeting cryptocurrency~\cite{conti2018survey}. 
The proliferation of phishing scams can lead to doubts about the security of blockchain technology, thereby hindering the healthy development of the technology~\cite{bartoletti2020dissecting}. 
Therefore, researchers have paid more attention to financial security in blockchain, and have come up with various methods to detect phishing fraud.

The openness and transparency of the blockchain makes transaction records available, further providing conditions for phishing detection.
Inspired by network science, researchers construct transaction networks from raw transaction records and apply various network analysis methods to detect phishing scams. 
Existing phishing detection methods~\cite{wu2020phishers,chen2020phishing,wang2021tsgn} mainly utilize graph representation learning techniques to generate the account feature vectors, and further achieve phishing detection via downstream machine learning classifiers.
However, these methods do not achieve an end-to-end architecture, failing in learning the task-related features.
Meanwhile, these methods generally suffer from poor scalability in large-scale Ethereum transaction network. 

In this paper, we consider phishing detection as a graph classification task.
Specifically, we first introduce a network lightweight strategy to rescale the Ethereum datasets and construct the Ethereum transaction networks.
Then we formulate a sampling rule for subgraph extraction, which can efficiently extract subgraphs of similar scales to allow for mini-batch training of subsequent models.
Finally, we design an end-to-end Chebyshev graph convolutional network to automatically extract account transaction behavior features and achieve phishing detection.
We conduct extensive experiments on five Ethereum datasets, and the results demonstrate that our method significantly outperforms existing phishing detection methods and scale well in large transaction networks.

The remainder of this work is organized as follows. 
In Section~\ref{sec:RelatedWork}, we briefly summarize the phishing detection methods and the network representation learning methods. 
In Section~\ref{sec:Methodology}, we introduce the details of our proposed phishing detection method, including lightweight network construction, transaction subgraph sampling, and the architecture of phishing detection graph neural network~(PDGNN). 
In Section~\ref{sec:Experiment}, we present the experimental setup and analyze the experimental results. 
Finally, we summarize the main work and contribution of this paper in Section~\ref{sec:Conclusion}.

\section{Related Work} \label{sec:RelatedWork}
\subsection{Phishing Detection Method}
The most existing illegal phishing fraud detection methods concentrate on graph analytics, which rely on the transaction networks constructed from raw blockchain transaction data.
Yuan et al.~\cite{wu2020phishers} proposed a graph random walk method named trans2vec to extract the account features for phishing detection on Ethereum, which takes the transaction amount and timestamp into account and uses SVM classifier to distinguish the accounts into normal and phishing ones.
Chen et al.~\cite{chen2020phishing} proposed a phishing detection method based on graph convolutional network (GCN) and autoencoder, which aggregates account features and network topology, and uses the lightGBM classifier to identify phishing accounts in Ethereum.
Wang et al.~\cite{wang2021tsgn} introduced the subgraph network (SGN) mechanism into Ethereum transaction network for extending the feature space of accounts. They constructed the transaction subgraph networks (TSGN) by extracting the first-order subgraphs of accounts, and used graph neural networks such as GCN and Diffpool to identify phishing accounts.
Shen et al.~\cite{shen2021identity} proposed a generic end-to-end graph neural network model named $\mathrm{I^{2}BGNN}$, which can accept transaction subgraphs as input and learn to map the subgraph patterns to the label associated with account identity.
Zhou et al.~\cite{zhou2022behavior} proposed a joint learning graph neural network framework named Ethident, which utilizes a hierarchical graph attention encoder to characterize the account behavior pattern and applies self-supervision mechanism for improving model generalization.

\subsection{Network Representation Learning}
Network representation Learning aims to learn latent, low-dimensional representations of network nodes, while preserving network topology structure, node content, and other side information~\cite{zhang2018network}. 
The generated low-dimensional node features can be easily used for downstream graph analysis tasks, such as node classification~\cite{abu2020n}, graph classification~\cite{xuan2019subgraph}, link prediction~\cite{chen2021time}, etc. 
Perozzi et al.~\cite{perozzi2014deepwalk} proposed to generate a node representation through the co-occurrence probability of the node sequence generated by random walk. 
Grover et al.~\cite{grover2016node2vec} proposed biased random walks to balance Breadth First and Depth First sampling, achieving a balance between homogeneity and structural equivalence.
The factorization-based graph embedding algorithm uses the connection information between nodes to construct various matrices~(such as Laplace matrix and adjacency matrix) and then decomposes the above matrix to obtain node embedding vectors. 
The models related to matrix decomposition include Graph Factorization~(GF)~\cite{ahmed2013distributed}, GraRep~\cite{cao2015grarep}, HOPE~\cite{ou2016asymmetric}, etc. 
Different from the above methods, LINE~\cite{tang2015line} optimizes the first-order and second-order proximity of nodes by designing a specific objective function to obtain node representation. 
The graph convolutional network (GCN) proposed by Kipf et al.~\cite{kipf2016semi} iteratively aggregates and updates the node features based on network topology, achieving higher expression performance.
The Graph2vec proposed by Narayanan et al.~\cite{narayanan2017graph2vec} can learn the graph-level representations of networks and serves for graph classification tasks.


\section{Methodology} \label{sec:Methodology}

\begin{figure}[!t]
  \centering
  \includegraphics[width=1\linewidth]{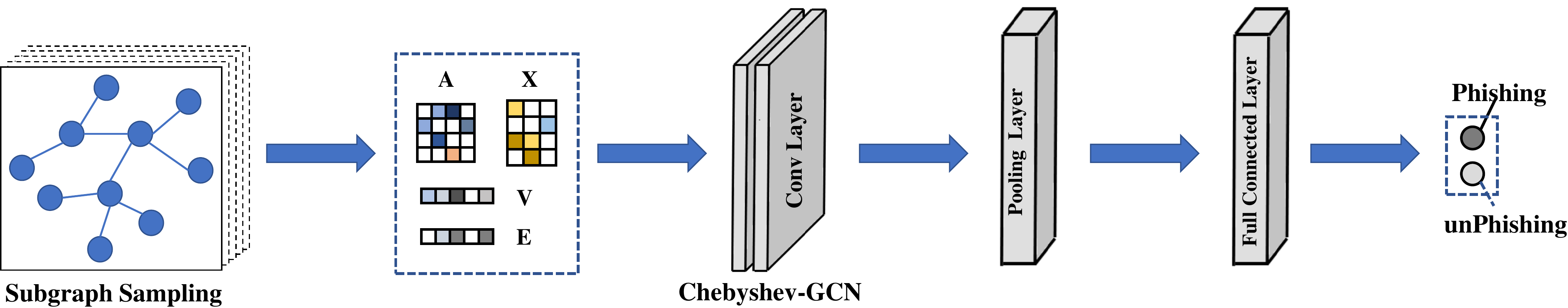}
  \caption{The overview of PDGNN framework for Ethereum phishing detection.}
  \label{fig:frame}
\end{figure}


\subsection{Problem Description}\label{sec:definition}
Ethereum transaction data can be modeled as a transaction network $G=(V,E,\textbf{X})$, where $V=\{v_{1},v_{2},\cdots,v_{n}\}$ represents the account set, $E=\{(v_{i},v_{j}) \mid i\neq j\}$ represents the set of transactions with timestamp and amount value, $\textbf{X}\subseteq \mathbb{R}^{n\times d}$ is the node feature matrix constructed from the information of transaction frequence. 
We consider phishing detection as a graph classification task.
Given a set of account subgraphs $\Omega =\{G_{i}, y_i\}$, graph classification aims to learn a function mapping the subgraph patterns $G_i$ to label $y_i$ associated with account identity, i.e., whether the target account is a phishing account or not.

\subsection{Lightweight Transaction Network}\label{sec:lightweight}

Thanks to the openness of blockchain, researchers can crawl transactions from Etherscan platform\footnote{\url{https://etherscan.io/}}. 
However, there are extremely large transaction records and most accounts are not always active, which makes it difficult to mine account characteristics through a large number of transactions. 
It is imperative to construct a lightweight network to reduce the complexity of phishing detection tasks. 
In this subsection, we focus on the labeled target accounts and extract transactions associated with them by a second-order Breadth First Search~(BFS). 
Firstly, we collect 1165 phishing accounts from Etherscan and perform a first-order BFS centered on these target accounts to extract transactions. 
After this, the total transaction information reduce to 1,686,003 accounts, 4,380,616 transaction records and 167 weakly connected components. 
However, the transaction network constructed from the above information is still large-scale, which is not conducive to those methods with full-graph learning manner. 
This paper uses only the largest weakly connected component, with a total of 1,684,164 accounts and 4,378,716 transaction records, and further propose a random walk-based sampling algorithm to lightweight the transaction network, i.e., a second-order BFS. 
During the random walk, we firstly select an account as an initial node and specify the network scale. 
Then we choose one of the neighbor accounts of the current account and move forward until the number of accounts in the network reaches the specified scale. 
If the neighbor of the current account does not exist, we will randomly select an account from the visited accounts.

\subsection{Subgraph Sampling}
When extracting the transaction subgraph of the target account, it is reasonable to consider the number of hops and sampled neighbors per hop, because they control the scale of subgraphs. 
In this subsection, we propose a formula for calculating the number of sampling neighbors, which ranks each node's neighbors and samples its top-$k$ neighbors according to the edge information. 
The calculation formula is as follows: 
\begin{equation}\label{eq:1}
k=\lceil \overline D \times (1+Density) \rceil =\lceil \frac{\vert 2E\vert}{\vert V\vert} \times (1+\frac{2\vert E\vert }{\vert V\vert (\vert V\vert -1)}) \rceil
\end{equation}
where $\overline D$ is the average network degree, $Density$ is the network density, $\lceil \cdot \rceil$ represents ceil operation, $\vert V\vert$ and $\vert E\vert$ represent the number of nodes and edges respectively. 
The calculation formula adaptively extracts subgraphs of suitable scale according to the network topology, which improves the subsequent calculation efficiency compared with the traditional sampling strategy with fixed number of neighbors.
In addition, the transaction edges contain two attributes: transaction amount value $\mathbf{a}$ and times $\mathbf{t}$, so that the neighbors of the target account can be sorted separately according to these two attributes. 
When the number of neighbors is less than $k$, all neighbors will be extracted to construct subgraph.

\subsection{Phishing Detection based on Graph Neural Network}\label{sec:PDGNN}
In this section, we present the details of phishing scam detection graph neural network~(PDGNN), which is divided into the following steps: (1) automatically aggregate nodes' information by using graph convolutional network, (2) extract the characteristics of the target node by pooling layer, (3) classify phishing accounts through the full connection layer.

\subsubsection*{Graph Convolution Layer}\label{sec:GCN}
We use Chebyshev GCN to automatically aggregate and update account features, and the single-layer Chebyshev GCN is defined as:
\begin{equation}\label{eq:2}
  \textbf{H}=\sigma \left \{ \sum_{k=0}^{K-1}\beta_{k}T_{k}(\widetilde{ \textbf{L}})\textbf{X}\right \},
\end{equation}
where $\sigma (\cdot)$ is RELU activation function, $K$ is the number of layers, $\beta_{k}$ is the coefficient of Chebyshev polynomial, $\textbf{X}$ is the node feature matrix. 
$T_{k}(\widetilde {\textbf{L}})$ is $k$ order Chebyshev polynomial and $\widetilde {\textbf{L}}=2\textbf{L}/\lambda_{max}-\textbf{I}$ where $\lambda_{max}$ can be solved by power iteration, $\textbf{L}$ is the Laplace matrix. 
The transaction subgraph extracted in this paper is a weighted directed network, thus the Laplace matrix needs to be transformed as:
\begin{equation}\label{eq:3}
\textbf{L}=\textbf{I}-\widetilde {\textbf{D}}^{-\frac{1}{2}}\textbf{A}\widetilde {\textbf{D}}^{-\frac{1}{2}}, \ \widetilde {\textbf{A}}=\textbf{A}+\textbf{A}^\top ,
\end{equation}
where $\widetilde {\textbf{D}}$ is the normalized degree matrix of $\widetilde {\textbf{A}}$. 
$T_{k}(\widetilde {\textbf{L}})$ recursion can be carried out by using the properties of Chebyshev polynomial:
\begin{equation}\label{eq:4}
T_{k}(\widetilde {\textbf{L}})=2\widetilde {\textbf{L}} T_{k-1}(\widetilde {\textbf{L}})-T_{k-2}(\widetilde {\textbf{L}}), T_{0}(\widetilde {\textbf{L}})=\textbf{I}, T_{1}(\widetilde {\textbf{L}})=\widetilde {\textbf{L}} .
\end{equation}


\subsubsection*{Pooling Layer}\label{sec:Pooling}
Here we use the pooling layer to obtain the target account's representation from the whole graph: 
\begin{equation}\label{eq:5}
\textbf{Z}=\mathrm{pooling}(\textbf{H}),
\end{equation}
where $\mathrm{pooling}(\cdot)$ is a pooling function. 
Here, we adopt the average pooling function.

\subsubsection*{Full Connected Layer}\label{sec:full}
We use full connected layer for graph classification: 
\begin{equation}\label{eq:6}
\hat y=\mathrm{softmax}(\textbf{WZ}+\textbf{b}),
\end{equation}
where $\textbf{W}$ and $\textbf{b}$ are parameters of full connected layer. 
The model uses cross-entropy loss function:
\begin{equation}\label{eq:7}
\mathcal{L}=-\frac{1}{n} \sum_{i=1}^{N} y_i\log\hat y_i - (1-y_i)\log(1-\hat y_i).
\end{equation}

\begin{table*}[!t]
  \setlength{\tabcolsep}{6.5mm}
  \centering
  \renewcommand{\arraystretch}{1.2}
  \setlength{\abovecaptionskip}{0pt}
  \setlength{\belowcaptionskip}{5pt}
  \caption{Basic statistics of the Ethereum datasets. $|V|$ and $|E|$ are the numbers of nodes and edges respectively, $Y$ is the number of labeled phishing nodes.}
  \begin{tabular}{ccccc}
  \hline\hline
  Dataset     & $|V|$      & $|E|$    & $|Y|$   \\ 
  \hline
  EthereumG1  & 20000      & 131189     & 242               \\
  EthereumG2  & 30000      & 172011     & 363                 \\
  EthereumG3  & 40000      & 202595     & 462                  \\ 
  EthereumG4  & 50000      & 227854     & 556                  \\ 
  EthereumG5  & 60000      & 250402     & 604                  \\ 
  \hline\hline
  \end{tabular}
  \label{tab:dataset}
\end{table*}

\section{Experimental Results} \label{sec:Experiment}
\subsection{Dataset}
We evaluate the proposed method on five lightweight Ethereum networks. The basic statistics are summarized in Table~\ref{tab:dataset}.

\subsection{Comparison Methods and Experimental Setup}
We compare PDGNN with seven graph-based detection algorithms, including manual feature~(MF), LINE~\cite{tang2015line}, DeepWalk~(DW)~\cite{perozzi2014deepwalk}, Node2Vec~(N2V)~\cite{grover2016node2vec}, T-EDGE~\cite{lin2020t}, Graph2Vec~(G2V)~\cite{narayanan2017graph2vec} and $\mathrm{I^{2}BGNN}$~\cite{shen2021identity}. 
For all random walk-based methods including DW, N2V and T-EDGE, we use the same parameter settings: the number of walks per node is 10, the length of walks is 30 and the size of the context window is 5. 
For all graph embedding methods, the embedding dimension is set to 128. 
For manual feature method and all graph embedding methods, we use logistic regression classifier for classification. 
For the end-to-end GNN-based models $\mathrm{I^{2}BGNN}$ and PDGNN, the dimension of the two hidden layers is set to 128. 
We split each dataset into training and testing sets with a proportion of 4:1, we repeat 5-fold cross validation for 5 times and report the average performance measures involving Precision, Recall, F1-score and Accuracy.

\subsection{Evaluation Metrics}
For a binary node classification problem, the prediction results can be divided into the following four cases according to the real label and predicted label of the node:
\begin{itemize}
  \item[$\bullet$] \textbf{True positive (TP):} The real label is positive, and the predicted label is positive.
  \item[$\bullet$] \textbf{False positive (FP):} The real label is positive, but the predicted label is positive.
  \item[$\bullet$] \textbf{False negative (FN):} The real label is positive but the predicted label is negative.
  \item[$\bullet$] \textbf{True negative (TN)} The real label is negative, and the predicted label is negative.
\end{itemize}

Precision, recall, F1-score and Accuracy are most important indicators in node classification, and their definitions will be briefly introduced below: 
\begin{itemize}
  \item[$\bullet$] \textbf{Precision:} It shows how many of the samples with positive category are real positive samples and is defined as:
  \begin{equation*}
    \rm Precision = \frac{TP}{TP+FP}
  \end{equation*}
  
  \item[$\bullet$] \textbf{Recall:} Recall indicates how many positive examples in the sample are predicted correctly and is defined as:
   \begin{equation*}
    \rm Recall = \frac{TP}{TP+FN}
  \end{equation*}

   \item[$\bullet$] \textbf{F1-score:} F1-score is the harmonic average of Precision and Recall without being influenced by unbalanced samples. The calculation formula is: 
   \begin{equation*}
    \rm F1-score = 2 \frac{Recall\times Precision}{Recall + Precision}
  \end{equation*}

  \item[$\bullet$] \textbf{Accuracy:} Accuracy means that the correctly classified samples are divided by the total number of samples and is defined as :
   \begin{equation*}
    \rm Accuracy = \frac{TP+TN}{TP+FP+TN+FN}
  \end{equation*}
  
\end{itemize}

\begin{table}[!t]

  \setlength{\tabcolsep}{1.2mm}
  \centering
  \renewcommand{\arraystretch}{1.2}
  \setlength{\abovecaptionskip}{0pt}
  \setlength{\belowcaptionskip}{5pt}
  \caption{Results of phshing detection. The best results are highlighted in blod.}
  \resizebox{\textwidth}{!}{
  \begin{tabular}{c|ccccccccc}
  \hline \hline
  \multicolumn{10}{c}{EthereumG1}  \\
  \hline
  Algorithm &MF &LINE &DW &N2V &T-EDGE &G2V-$t$ &G2V-$a$ &$\mathrm{I^{2}BGNN}$ &PDGNN\\
  \hline
  Precision &0.6968 &0.6512 &0.8119 &0.7331 &0.7479 &0.7662 &0.7816 &0.7726 &\textbf{0.8319} \\
  Recall    &0.6161 &0.4173 &0.7769 &0.6651 &0.7186 &0.7640 &0.7398 &0.8963 &\textbf{0.9377} \\
  Accuracy  &0.5093 &0.5938 &0.7979 &0.7072 &0.7381 &0.7649 &0.7670 &0.8165 &\textbf{0.8742} \\
  F1-score  &0.4477 &0.5057 &0.7935 &0.6916 &0.7319 &0.7631 &0.7595 &0.8294 &\textbf{0.8813} \\
  \hline
  
  \multicolumn{10}{c}{EthereumG2}  \\
  \hline
  Algorithm &MF &LINE &DW &N2V &T-EDGE &G2V-$t$ &G2V-$a$ &$\mathrm{I^{2}BGNN}$ &PDGNN\\
  \hline
  Precision &0.8272 &0.6980 &0.8061 &0.8282 &0.8308 &0.8272 &0.8045 &0.8437 &\textbf{0.8875} \\
  Recall    &0.3890 &0.4795 &0.7041 &0.7014 &0.7890 &0.8164 &0.8082 &0.8904 &\textbf{0.9205} \\
  Accuracy  &0.6534 &0.6356 &0.7658 &0.7712 &0.8137 &0.8219 &0.8055 &0.8616 &\textbf{0.9014} \\
  F1-score  &0.5252 &0.5674 &0.7497 &0.7517 &0.8089 &0.8211 &0.8053 &0.8656 &\textbf{0.9033} \\
   \hline
  \multicolumn{10}{c}{EthereumG3}  \\
  \hline
  Algorithm &MF &LINE &DW &N2V &T-EDGE &G2V-$t$ &G2V-$a$ &$\mathrm{I^{2}BGNN}$ &PDGNN\\
  \hline
  Precision &0.6622 &0.6328 &0.8205 &0.8227 &0.8041 &0.8069 &0.8103 &0.8269 &\textbf{0.8551} \\
  Recall    &0.6409 &0.4501 &0.7921 &0.7402 &0.7533 &0.7985 &0.7490 &0.8699 &\textbf{0.9068} \\
  Accuracy  &0.5232 &0.5957 &0.8097 &0.7795 &0.7838 &0.8032 &0.7870 &0.8432 &\textbf{0.8768} \\
  F1-score  &0.4753 &0.5252 &0.8059 &0.7658 &0.7763 &0.8015 &0.7783 &0.8470 &\textbf{0.8800} \\
   \hline

  \multicolumn{10}{c}{EthereumG4}  \\
  \hline
  Algorithm &MF &LINE &DW &N2V &T-EDGE &G2V-$t$ &G2V-$a$ &$\mathrm{I^{2}BGNN}$ &PDGNN\\
  \hline
  Precision &0.6687 &0.6993 &0.8454 &0.8210 &0.8234 &0.8252 &0.8192 &0.8156 &\textbf{0.8648} \\
  Recall    &0.6017 &0.5439 &0.8044 &0.7720 &0.7972 &0.8455 &0.8563 &0.8707 &\textbf{0.9137} \\
  Accuracy  &0.5022 &0.6547 &0.8287 &0.7982 &0.8126 &0.8314 &0.8332 &0.8359 &\textbf{0.8852} \\
  F1-score  &0.4249 &0.6103 &0.8238 &0.7919 &0.8093 &0.8339 &0.8367 &0.8404 &\textbf{0.8883} \\
   \hline

\multicolumn{10}{c}{EthereumG5}  \\
  \hline
  Algorithm &MF &LINE &DW &N2V &T-EDGE &G2V-$t$ &G2V-$a$ &$\mathrm{I^{2}BGNN}$ &PDGNN\\
  \hline
  Precision &0.6509 &0.6765 &0.8015 &0.8184 &0.8240 &0.8160 &0.8079 &0.8331 &\textbf{0.8930} \\
  Recall    &0.8033 &0.5603 &0.7934 &0.7521 &0.7537 &0.8413 &0.8248 &0.8694 &\textbf{0.8992} \\
  Accuracy  &0.6496 &0.6463 &0.7975 &0.7860 &0.7959 &0.8256 &0.8140 &0.8463 &\textbf{0.8950} \\
  F1-score  &0.6966 &0.6126 &0.7964 &0.7755 &0.7867 &0.8284 &0.8160 &0.8502 &\textbf{0.8958} \\
   \hline \hline

  \end{tabular}
  }
  \label{tab:lp1}%
\end{table}

\subsection{Evaluation on Phishing Detection}
We compare our PDGNN with seven baseline methods to evaluate its effectiveness in detecting phishing accounts, and the experimental results are reported in Table~\ref{tab:lp1}, from which we observe that our PDGNN achieve state-of-the-art results with respect to comparison methods.

Across all datasets, the manual feature method~(MF) has the worst detection performance, because simple heuristic features are not capable of capturing structural pattern features of accounts. 
Compared with manual features, random walk-based methods (DW, N2V, T-EDGE) achieve better performance in detecting phishing accounts, because they can learn the structural pattern information of accounts.
Graph2Vec and $\mathrm{I^{2}BGNN}$ consider phishing detection as a graph classification task, and outperform other baselines with node classification manner, indicating that the key behavior pattern information of accounts can be preserved in the account subgraphs.

For our PDGNN, it significantly outperforms manual feature method (MF) across all datasets, and yields 28.60\% $\sim$ 109.21\% relative improvement in terms of F1-score, indicating that the learned account features are better at characterizing the behavior patterns of phishing accounts than the heuristic features.
When compared to graph embedding methods, our PDGNN surpasses strong baselines: we observe 6.17\% $\sim$ 11.06\% relative improvement over best baselines.
Our PDGNN uses structural information to update the attribute features of accounts and optimizes them in an end-to-end manner, which significantly outperforms graph embedding methods that only extract structural information of account interactions and detect phishing accounts in a two-stage manner.
We also compare with end-to-end GNN-based method $\mathrm{I^{2}BGNN}$, and observe 3.90\% $\sim$ 6.26\% relative improvement, which may benefit from the better expression ability of Chebyshev GCN.

\subsection{Parameter Analysis} 
We further investigate the impact of different parameters in our PDGNN.

\subsubsection{Impact of Sampling Strategies}
\begin{figure}[!t]
  \centering
  \includegraphics[width=0.7\linewidth]{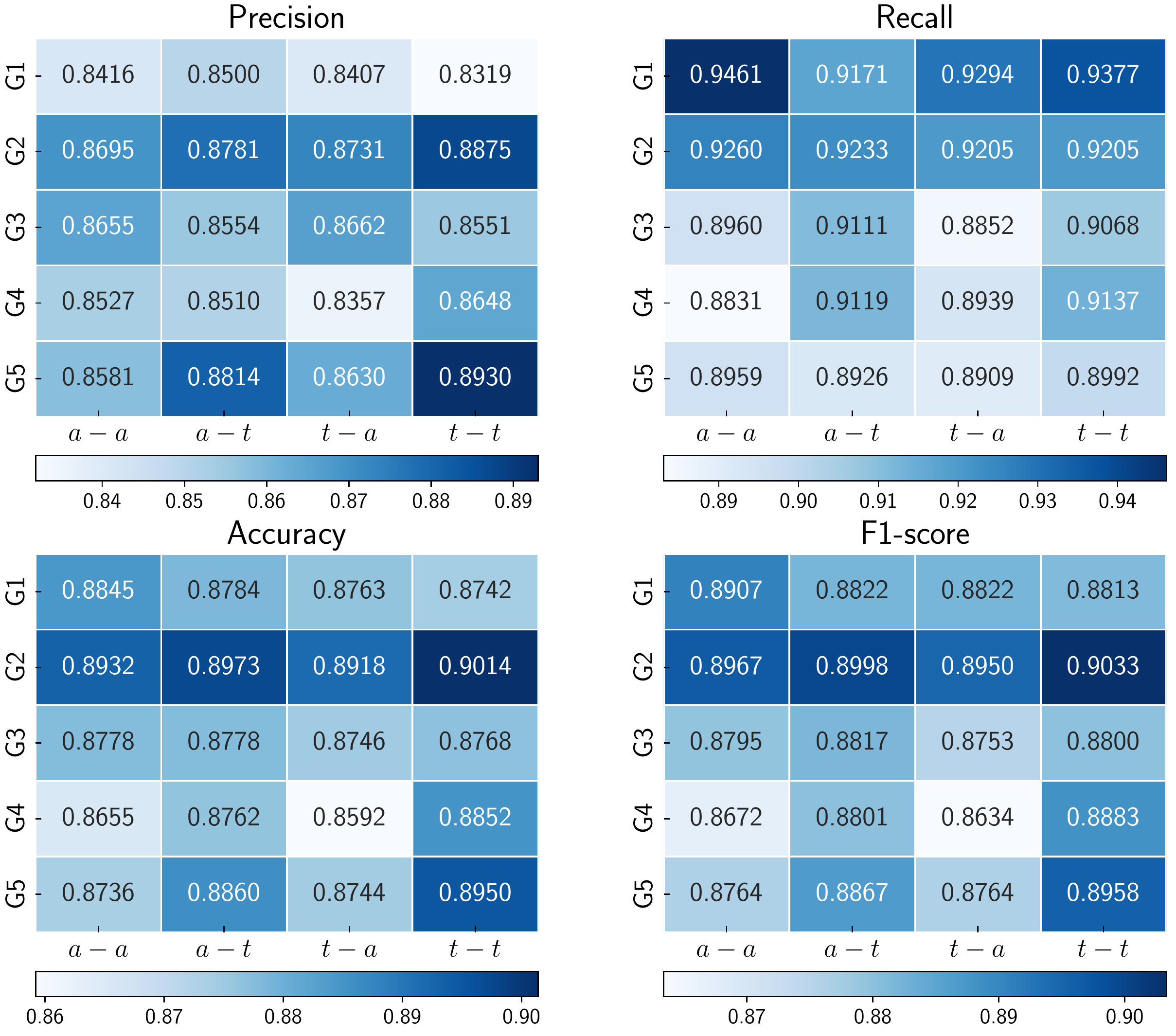}
  \caption{Phishing detection results under different sampling strategies.}
  \label{fig:sampling}
\end{figure}

We sample transaction subgraphs according to different edge information, and preserve one of the edge information, yielding four types of subgraph datasets: $\mathbf{a}-\mathbf{a}$, $\mathbf{a}-\mathbf{t}$, $\mathbf{t}-\mathbf{a}$ and $\mathbf{t}-\mathbf{t}$.
For example, $\mathbf{a}-\mathbf{t}$ represent that we sample transaction subgraphs according to the transaction amount value $\mathbf{a}$ and preserve the transaction times $\mathbf{t}$ as edge weight.
Table~\ref{fig:sampling} shows the phishing detection results under different sampling strategies, from which we observe that $\mathbf{a}-\mathbf{t}$ and $\mathbf{t}-\mathbf{t}$ achieve better performance in most cases.
We have reasonable explanations for such phenomenon.
Phishers engage in illegal fraud activities, they usually spread a large number of websites, emails or links containing viruses, unwanted software, etc., and trick the
recipient into doing remittances directly. 
As a result, the center phishing account would receive large number of transactions from those victims, which is significantly different from the behavior patterns of normal accounts.
Thus we speculate that the transaction times $\mathbf{t}$ (i.e., the number of transactions) would play an important role in distinguishing phishing accounts.

\subsubsection{Impact of Directivity}
\begin{figure}[!t]
  \centering
  \includegraphics[width=0.7\linewidth]{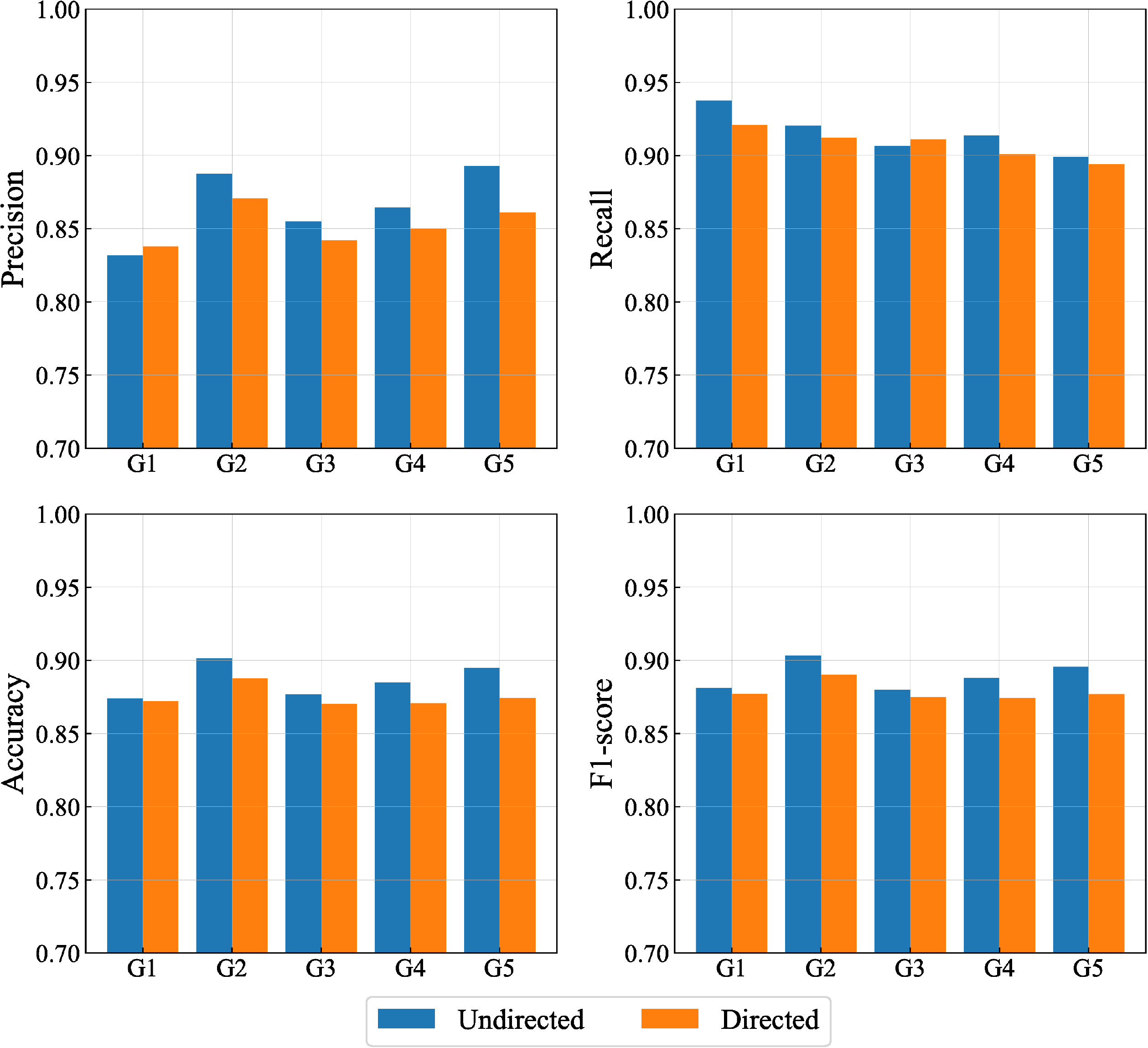}
  \caption{Phishing detection results under directed and undirected networks.}
  \label{fig:directed}
\end{figure}

Here, we investigate the impact of the network directivity on the proposed PDGNN, as shown in Fig.~\ref{fig:directed}. 
As we can see, training with undirected networks seems to benefit the model generalization more, indicating that undirected networks greatly simplify the complex and redundant structure in the Ethereum transaction network and have better detection performance than directed ones.

\begin{figure}[!t]
  \centering
  \includegraphics[width=0.7\linewidth]{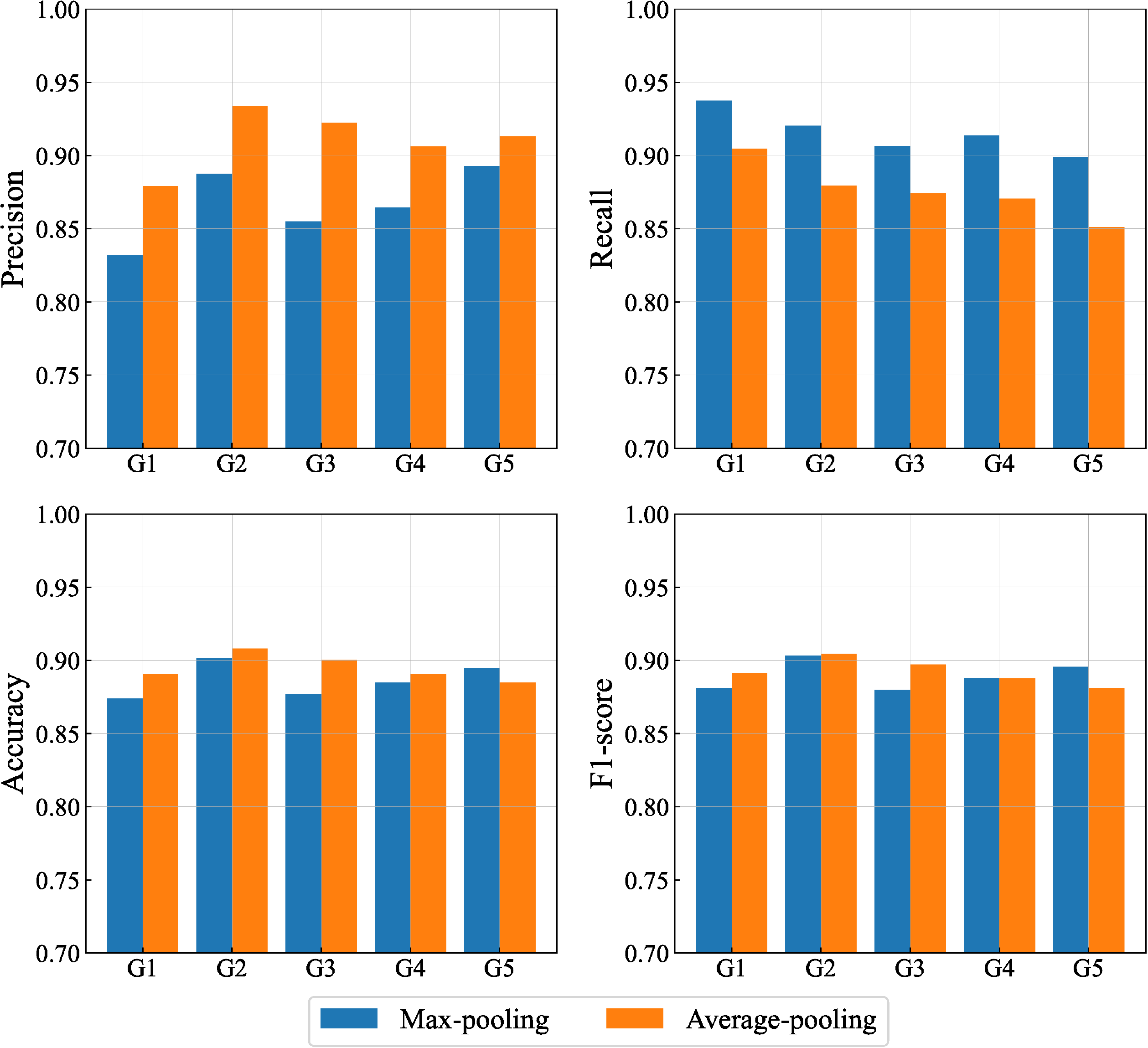}
  \caption{Phishing detection results under different graph pooling functions.}
  \label{fig:pooling}
\end{figure}

\subsubsection{Impact of Graph Pooling}
Here we use max-pooling and average-pooling to test the performance of subgraph representation.
Max-pooling can extract the most important features of each node in the subgraph, while average-pooling can keep the information of the whole subgraph and make features smoother.
Fig.~\ref{fig:pooling} shows the phishing detection results under different graph pooling functions, from which we observe that average-pooling outperforms max-pooling in most cases, indicating that the characteristics of all nodes play an important role in representing the behavior patterns reflected in subgraph.

\subsubsection{Impact of Hidden Dimension}
\begin{figure}[!t]
  \centering
  \includegraphics[width=0.7\linewidth]{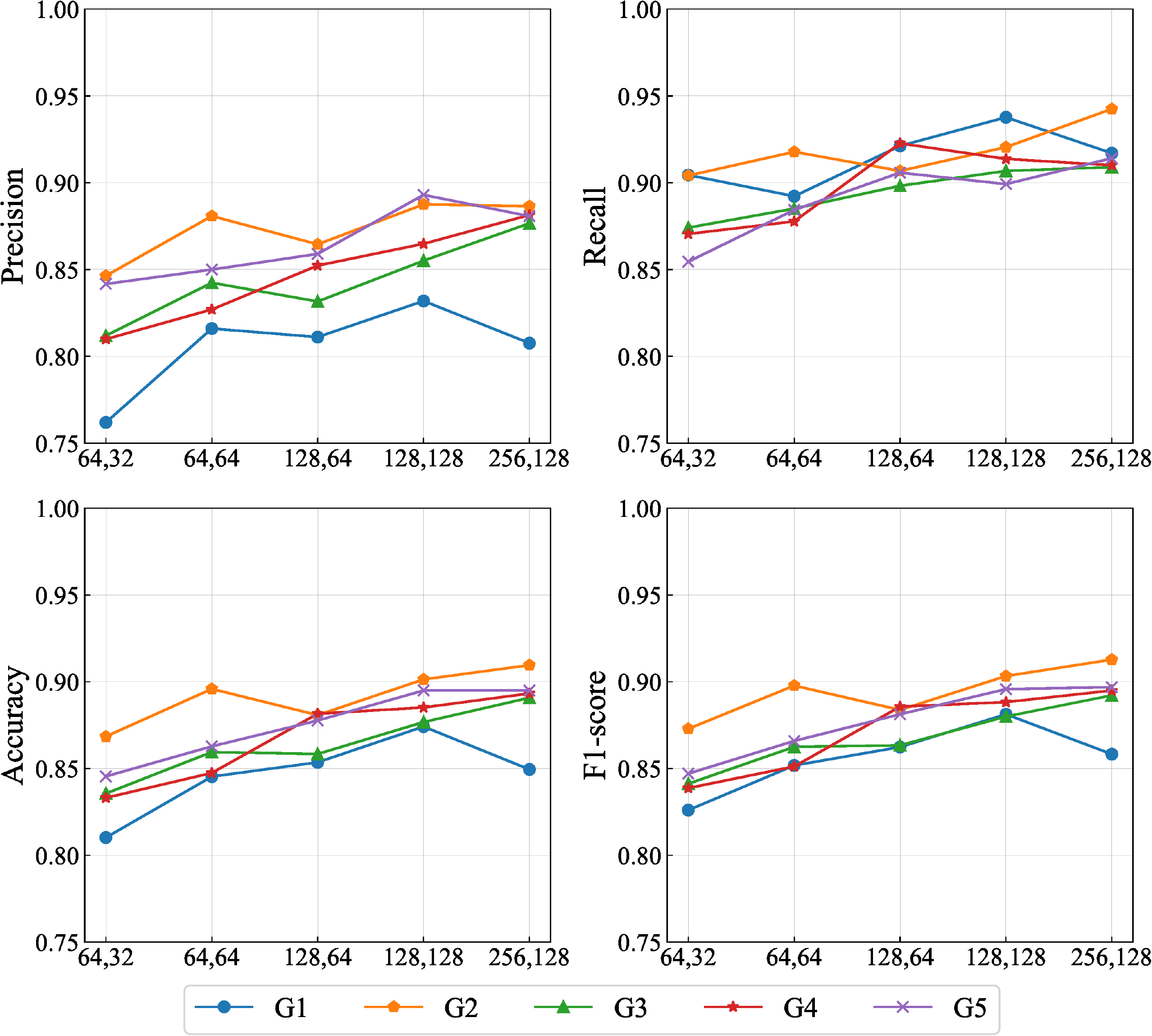}
  \caption{Phishing account detection results under different dimensions of the convolutional layer.}
  \label{fig:dimension}
\end{figure}

GNN is usually applied to shallow networks, and simply stacking more layers will lead to feature over-smoothing. 
In this paper, we apply two GNN layers in PDGNN. 
The impact of hidden dimensions is shown in Fig.~\ref{fig:dimension}, from which we can observe that the phishing detection performance is proportional to the dimension of the hidden layer to a certain extent, i.e., high dimensional embedding can encode more information. 
However, when the hidden dimension increases continuously, the performance tends to decrease. 
The reason is that the model with high dimension GNN layer will be overfitting, resulting in poor model generalization.

\section{Conclusions and Future Work} \label{sec:Conclusion}
In this paper, we propose a deep model phishing detection framework based on graph classification, which provides a novel network perspective of to detects phishing scam. Specifically, we present a lightweight network method to rescale dataset and make subgraph sampling to reduce in resource consumption. Further, we designi an end-to-end PDGNN model to learn user behaviors and detect phishing accounts. Experimental results demonstrate the effectiveness of our proposed phishing detection framework, and indicate that PDGNN is capable of extracting more information from Ethereum transactions. 

For future work, we plan to design more comprehensive subgraph sampling rules to make full use of the Ethereum transaction information. At the same time, we hope analyze account behavior from more application scenarios to create a secure transaction environment.

\subsubsection*{Acknowledgments} 
This work was partially supported by the National Key R\&D Program of China under Grant 2020YFB1006104, by the Key R\&D Programs of Zhejiang under Grant 2022C01018, by the National Natural Science Foundation of China under Grant 61973273, and by the Zhejiang Provincial Natural Science Foundation of China under Grant LR19F030001.

\bibliography{ref} 
\bibliographystyle{splncs04_}

\end{document}